\documentclass[10pt,aps,prb,raggedbottom,longbibliography,reprint,citeautoscript,letterpaper]{revtex4-2}
\usepackage[usenames,dvipsnames]{color}
\usepackage{graphicx,microtype}
\usepackage{multirow} 
\usepackage[bookmarks=false,colorlinks]{hyperref}
\hypersetup{linkcolor=magenta,citecolor=MidnightBlue,filecolor=Plum,urlcolor=MidnightBlue}
\usepackage[all]{hypcap} 
\usepackage{amsmath}
\usepackage{gensymb}
\usepackage{comment}
\usepackage{enumitem,amssymb}
\newlist{todolist}{itemize}{2}
\setlist[todolist]{label=$\square$}
\usepackage{pifont}



\usepackage{soul}

\begin{document}

\title{Anion correlation induced nonrelativistic spin splitting in rutile antiferromagnets
} 
\author{Siddhartha S.\ Nathan}
\affiliation{Department of Materials Science and Engineering, Northwestern University, Evanston, Illinois  60208, USA}

\author{Danilo Puggioni}
\affiliation{Department of Materials Science and Engineering, Northwestern University, Evanston, Illinois  60208, USA}

\author{Linding Yuan}
\email{linding.yuan@northwestern.edu}
\affiliation{Department of Materials Science and Engineering, Northwestern University, Evanston, Illinois  60208, USA}

\author{James M.\ Rondinelli}
\email{jrondinelli@northwestern.edu}
\affiliation{Department of Materials Science and Engineering, Northwestern University, Evanston, Illinois  60208, USA}

\begin{abstract}
Many studies of non-relativistic spin-splitting (NRSS), or altermagnetism, have focused on idealized, perfectly ordered crystals, relying on symmetry-based approaches to identify candidate materials. 
Here, we theoretically investigate how local short-range ordering (SRO) influences NRSS of energy bands in partially ordered collinear antiferromagnetic iron oxyfluoride (FeOF). 
Using the cluster expansion method, we identify four nearly degenerate structures (energy difference $\leq 8$ meV per formula unit) that represent distinct snapshots of local plane-to-plane O/F correlations.
Our density functional theory (DFT) results show robust NRSS along the $\Gamma$-M direction in all four structures, despite the absence of long range order. 
The magnitude and character of the splitting depend sensitively on the specific direction of anion correlations, effects that are not fully captured in high-symmetry average structures. 
Notably, two configurations ($Pmc2_1$ and $Pm$) exhibit $\Gamma$-point spin splitting absent in ordered FeF$_2$ and a virtual crystal approximation model of FeOF. 
We further predict distinct magneto-optical Kerr effect (MOKE) signatures, enabling experimental detection of SRO-driven electronic structure changes.
These results highlight heteroanionic compounds as a promising design space for NRSS antiferromagnets, with  experimentally synthesized FeOF already exhibiting a substantially higher N\'eel temperature (315\,K) than FeF$_2$ (79\,K).
\end{abstract}

\date{\today}

\maketitle

\section{Introduction}
\begin{figure*}
    \includegraphics[width=0.94\textwidth]{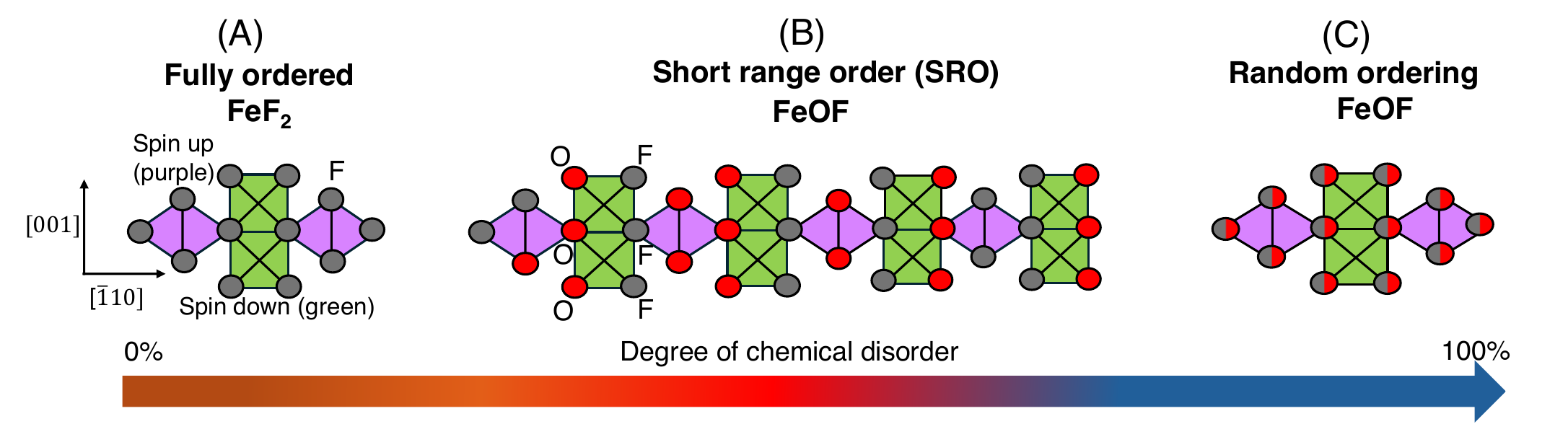}
    \caption{Schematic illustration of (a) fully ordered FeF$_2$ and (c) fully disordered FeOF structure in the rutile framework. (b) Anionic short-range order (SRO) in FeOF), illustrating the ordering within the (110) plane with an  $\cdots$O-F/O-F$\cdots$ pattern along [001] but with variable plane-to-plane correlations. Magnetic ordering of spin-up and spin-down sublattices are color coded as purple and green in their spin structure motif pairs (SSMPs), respectively.}
    \label{fig:intro}
\end{figure*}

Collinear antiferromagnets (AFMs)  composed of elements with unpaired $d$ or $f$ electrons can, under certain symmetry conditions, exhibit spin-polarized energy-split bands even in the absence of relativistic spin-orbit coupling/
This phenomenon, known as non-relativistic spin splitting (NRSS) \cite{noda2016momentum, hayami2019momentum,naka2019spin,yuan2020giant, hayami2020bottom,vsmejkal2020crystal,yuan2021prediction,naka2021perovskite,yuan2021strong,egorov2021colossal} defines a distinct class of materials termed altermagnets \cite{vsmejkal2022beyond,PhysRevX.12.040501}.
Remarkably, NRSS in altermagnets can be one to two orders of magnitude larger than Rashba or Dresselhaus spin splitting in nonmagnetic materials \cite{rashba2015symmetry,dresselhaus1955spin}, making it highly promising for antiferromagnetic spintronics  where robust spin polarization coexists with zero net magnetization \cite{naka2019spin,naka2021perovskite,gonzalez2021efficient,vsmejkal2022giant,vsmejkal2020crystal}.
In the most extensively studied altermagnets, \emph{i.e.}, rutile-structured transition metal oxides and fluorides, such as RuO$_2$, MnF$_2$, and CoF$_2$,
the magnetic unit cell consists of two corner-sharing octahedra which define a spin structure motif pair (SSMP). Each unit features a transition metal (M) center surrounded by six anions (X), with semilocal magnetic moments on the metal sites aligned antiparallel (\autoref{fig:intro}a-c). 
This arrangement leads to distinct local crystal fields and a hidden ferroically ordered magnetic octupoles \cite{PhysRevX.14.011019} on the two magnetic sublattices, thus enable momentum-dependent spin splitting \cite{yuan2023degeneracy}. 

Most studies \cite{song2025altermagnets,wei2024crystal, fender2025altermagnetism} on NRSS to date have focused on idealized, perfectly ordered crystals (\autoref{fig:intro}a), using symmetry-based approaches to identify candidate materials \cite{wei2024crystal,naka2019spin,yuan2020giant,yuan2021prediction,vsmejkal2022beyond}.
In realistic materials, atomic site disorder, whether intentional (e.g., doping) or unintentional (e.g., vacancies, chemical heterogeneity), can significantly alter these local crystal fields. When disorder is correlated across multiple unit cells short-range order (SRO) emerges. The effects of such SRO on NRSS remain poorly understood.

Rutile-type mixed-anion compounds  offer a compelling platform to investigate this effect.
In rutile oxyhalides, the anion sublattice is partially occupied by both O and X (halide) anions, forming [MO$_3$X$_3$] heteroleptic polyhedra where each cation is coordinated by two types of anions (\autoref{fig:intro}b). Although the average crystal structure retains the rutile framework, substitutional disorder removes the ideal periodicity, introducing local symmetry breaking and strain due to differences in ionic size and electronegativity. 
Such local structural disorder can significantly alter orbital overlaps and introduce fluctuations in exchange interactions $J$ through atomic displacements, thereby impacting magnetism and spin splitting. Importantly, heteroleptic coordination is not random in oxyhalides; short-range order (SRO) often emerges due to bonding preferences, particularly within the \{110\} planes along [001], and has been a subject of study in these systems both theoretically and in experiment \cite{brink2000electron,PhysRevMaterials.8.054602}.

Here, we theoretically investigate how short-range anion ordering influences the NRSS of energy bands in collinear antiferromagnets, using iron oxyfluoride FeOF with observed SRO \cite{brink2000electron} as a representative material.
By combining a cluster expansion model (CEM) with density functional theory (DFT) calculations, we simulate SRO effects in FeOF.
We find robust momentum-dependent spin splitting along $\langle110\rangle$ $k$-directions, closely resembling the NRSS features of long-range-ordered FeF$_2$, despite the absence of long-range order.
We show the magnitude and nature of the spin splitting (spin split at $\Gamma$ or not) depend sensitively on the direction of the anion correlations, effects that are not possible in long-range ordered FeF$_2$ or the virtual crystal approximation (VCA) model of FeOF, where anion disorder is treated as an average occupancy of the anion site in a higher symmetry structure (\autoref{fig:intro}c). 
We further propose magneto-optical Kerr effect (MOKE) measurements as a viable experimental probe for detecting the SRO-induced electronic structure changes.
The findings highlight the promise of heteroanionic compounds as a broader design space for NRSS antiferromagnets, with FeOF already offering a higher Néel temperature (315\,K) \cite{chappert1966effet,chappert1966structure} than FeF$_2$ (79\,K) \cite{goodenough1963magnetism,CHIRWA1980457}.

\section{Methods}

\subsection{Density Functional Theory}
DFT calculations were performed using the VASP \cite{kresse1996efficient} code. We used the projector augmented wave (PAW) technique \cite{kresse1999ultrasoft} with a plane-wave energy cutoff of 650 eV and the revised Perdew, Burke, and Ernzerhof for solids (PBESol) generalized gradient (GGA) exchange-correlation functional \cite{perdew2008restoring}. All calculations were performed using dense well-converged $\Gamma$-centered $k$ meshes of at least 4000 $k$ points per reciprocal atom, and each structure was relaxed (volume, cell shape and atomic positions unconstrained) until the forces on the atoms were less than 10$^{-3}$ eV \AA $^{-1}$ . The atomic simulation environment \texttt{(ase)} \cite{larsen2017atomic}, \texttt{sumo} \cite{ganose2018sumo} softwares were used to aid all calculations and for post-processing. 

To compare band structures of our FeOF SRO configurations to a consistent baseline, we used the band unfolding technique \cite{popescu2010effective,ku2010unfolding}, using the \texttt{vaspkit} \cite{wang2021vaspkit} tool. In this approach, the eigenstates obtained from a DFT calculation for a supercell are projected onto the Bloch states of the corresponding primitive cell. The resulting spectral weights or unfolding weights at each $k$-point quantify the contribution of each supercell eigenstate to a given primitive-cell at the corresponding $k$-point, allowing the reconstruction of an effective band structure in the primitive Brillouin zone (BZ). In this case, all band structures of our FeOF SRO configurations were unfolded onto the BZ of the rutile FeF$_2$ cell.

\subsection{Cluster Expansion Model}
To capture the dependence of relative O/F anion ordering on the energy landscape within the rutile FeF$_2$–FeO$_2$ phase space, we constructed a cluster expansion model (CEM). The CEM was parameterized by fitting to DFT-calculated energies for 238 symmetrically distinct Fe–(O/F) supercells, spanning compositions from 100\,\% O to 100\,\% F. 
The rutile FeF$_2$ structure (ICSD coll.\ code 9166) was used as a prototype to generate these structures, given the similarities in the lattice parameters of FeF$_2$ and FeOF \cite{brink2000electron}.
Supercells up to dimensions of $1\times 1 \times 3$, $1\times 3 \times1$, and $3 \times 1 \times 1$ were generated using the Integrated Cluster Expansion Toolkit \texttt{(icet)} Python module \cite{aangqvist2019icet}.
The CEM was also constructed and fitted using the \texttt{icet} module. 
Each structure was parametrized into symmetrically unique 2- and 3-body clusters, whose occupancies were fitted to DFT-calculated energies using LASSO (least absolute shrinkage and selection operator) regression \cite{tibshirani1996regression}. 

For our training set,  we considered different magnetic configurations for our DFT relaxations. For each structure, the Fe atoms were initialized with ferromagnetic and antiferromagnetic spin configurations, along with non-spin polarized initializations. Ultimately, given the difficulty in stabilizing most of our structures when considering spin polarization, we used non-spin polarized DFT relaxations to train our CEM. Although the energy scales obtained are likely to differ from the true ground states (when considering spin-polarized calculations), we do not expect the presence of magnetism to change the relative energy trends, resulting in a reasonably accurate CEM. 
Our non-magnetic cluster expansion training resulted in a model with 11 non-zero parameters (including 8 pairs and 1 triplet) and a Cross-Validation (CV) score of 0.013 eV per site, comparable to previous CEMs for HAMs.  

\begin{figure*}
    \centering
    \includegraphics[width=0.97\textwidth]{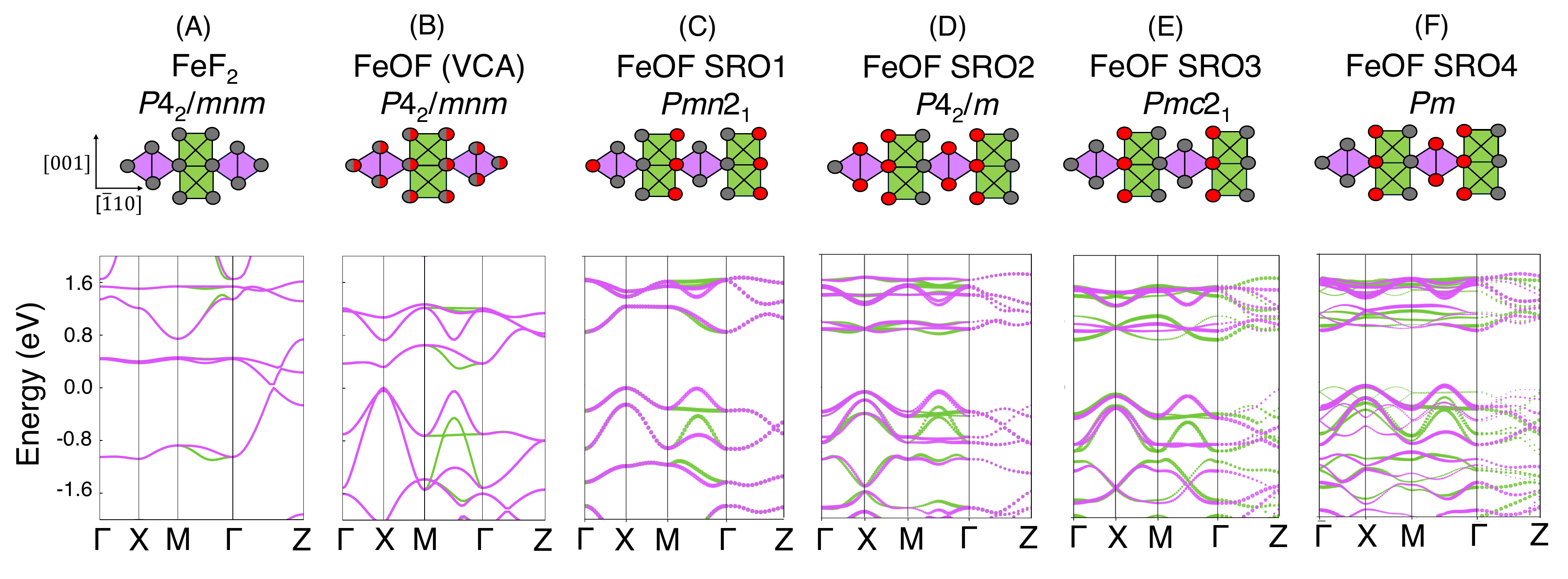}
    \caption{Crystal and electronic structures of (a) FeF$_2$, (b) VCA FeOF, and (c-f) four chosen FeOF configurations. The crystal structures shown are viewed along the rutile $c$ [001] axis (top-down view), showcasing the variations in O/F correlations. Their band structures are unfolded onto the primitive rutile unit cell for easy comparison, and the line widths represent the spectral weights at each $k$ point. The spin-up (magenta) and spin-down (cyan) motifs are connected by a 90\degree four-fold rotation in the (a) FeF$_2$ ($P4_2/mnm$) structure, (b) FeOF VCA ($P4_2/mnm$), (c) FeOF SRO1 ($Pmn2_1$), and (d) FeOF SRO2 ($P4_2/m$) cells, but not in the (e) FeOF SRO3 ($Pmc2_1$) and (f) FeOF SRO4 ($Pm$) cells, leading to spin-splitting of energy bands at $\Gamma$. All 4 structures show robust spin splitting along the $\Gamma-\mathrm{M}$ direction, which are contained in the (110) plane. Notice that the degeneracy of the energy bands at $\Gamma$ in panels (c,d) and the $\Gamma$ split bands in panels (e,f).}
    \label{fig:fef2_vca_plus_sro}
\end{figure*}

Following the construction of the CEM, we performed canonical Monte Carlo (MC) simulated annealing simulations on finite-sized supercells of FeOF. We used different sized supercells, ranging from $2\times 2 \times 2$ to $20\times 20 \times 20$. An exponential cooling function was chosen, and our system was allowed to anneal from 8000\,K to 0\,K for 400 MC sweeps (timesteps per atom).

\subsection{MOKE Simulations}
To calculate the MOKE spectrum for all structures, we used the independent particle approximation of the Green-Kubo formula \cite{gajdovs2006linear} to calculate the frequency dependent dielectric tensor $\varepsilon$  using VASP. 
The Kerr rotation spectrum was calculated using: 
\begin{equation*}
\theta_K(\omega) + i\,\epsilon_K(\omega) \;=\;
\frac{-\,\varepsilon_{xy}(\omega)}
{\bigl(\varepsilon_{xx}(\omega)-1\bigr)\,\sqrt{\varepsilon_{xx}(\omega)}} 
\end{equation*}
where $\theta_K$ is the real part of the expression and is used to describe the Kerr rotation. Naturally, the principal and off-diagonal terms can be modified according to the direction of the magnetization axis. Usually, for normal incidence along $z$, $\epsilon_{xx}$ is usually taken to be the average of the $\epsilon_{xx}$ and $\epsilon_{yy}$ terms. All calculations include SOC.

We note that one must be careful and use only the antisymmetric part of the off-diagonal component of the dielectric tensor to extract the Kerr spectrum. Since VASP reports 6 coefficients of the dielectric matrix, the procedure to extract the antisymmetric component is  as follows: (1) Perform the same independent approximation calculation with the magnetizations of all Fe atoms reversed along the easy axis. (2) Subtract the off-diagonal coefficients of $\epsilon(+M)$ and $\epsilon(-M)$ and halve this quantity to extract the anti-symmetric part, \emph{i.e.},
\[
    A_{ij} = \frac{\varepsilon_{ij}(+M)-\varepsilon_{ij}(-M)}{2}, i \neq j\,.
\]
The principal components $\varepsilon_{ii}, i=1,2,3$ should remain unchanged for both $+M$ and $-M$ calculations.

\section{Results}

\subsection{NRSS in long range ordered FeF$_2$}
FeF$_2$ has a centrosymmetric rutile structure (space group $P4_2/mnm$). The magnetic Fe$^{2+}$ ions are located at (0,\,0,\,0) and (1/2,\,1/2,\,1/2), and each is surrounded by an octahedra of non-magnetic F$^-$ anions.
The Fe moments order in a G-type antiferromagnetic (AFM) arrangement  aligned along the tetragonal $c$ axis (\autoref{fig:mag_order}a), corresponding to the magnetic space group $P4_2'/mnm'$. 
Following the Goodenough-Kanomori rules \cite{kanamori1959superexchange}, the exchange coupling along $c$ is ferromagnetic owing to 90$\degree$ exchange through the $e_{g}-p_{\sigma}-e_{g}$ channel. In contrast, the diagonal directions are antiferromagnetically coupled through 130$\degree$ superexchange involving both $e_{g}-p_{\sigma}-e_{g}$ and $t_{2g}-p_{\pi}-t_{2g}$. 
Our DFT simulations confirm that the G-AFM configuration is the ground state of FeF$_2$. The computed magnetic moment on Fe$^{2+}$ is 3.59\,$\mu_B$, which is in excellent agreement with the experimentally reported value \cite{strempfer2004magnetic}. 
\autoref{fig:fef2_vca_plus_sro}a presents the electronic band structure calculated along high-symmetry trajectories for the ordered rutile FeF$_2$ without spin-orbit coupling (SOC).
We find FeF$_2$ has a band gap of 0.35 eV (\autoref{fig:dos_fef2_vca}a), in good agreement with other first-principles approaches using the GGA method \cite{riss2003theoretical}.
Despite the relatively low atomic numbers of Fe ($Z=26$) and F ($Z=9$) in FeF$_2$, our calculations reveal robust spin  splitting along the $\Gamma$(0,0,0) to M($\pi$,$\pi$,0) path. The spin splitting of bands near the Fermi level, i.e., those most relevant for potential transport applications via doping, is on the order of 159\,meV.
The NRSS in FeF$_2$ arises from the combined effects of translation and inversion acting on non-interconvertible SSMPs, which break both $\Theta I$ and $UT$ symmetries \cite{yuan2021prediction,yuan2020giant}. 
These symmetries are essential for protecting spin degeneracy (see Appendix~\ref{sec:symm} for a detailed discussion). 
Here, $U$ denotes a spin rotation within the SU(2) group acting on the spin-$\frac{1}{2}$ space, effectively reversing the collinear spin; $T$ represents a spatial translation; $\Theta$ is the time-reversal operator; and $I$ corresponds to spatial inversion. 
In Litvin’s notation \cite{litvin1977spin}, these operations map to $\{C_2||E|t\}$ and $\{-1||-1\}$, consistent with the spin-symmetry framework \cite{PhysRevX.12.021016,vsmejkal2022beyond}.
In FeF$_2$, the following crystal rotation  and mirror operations $\{C_4|\frac{1}{2},\frac{1}{2},\frac{1}{2}\}$, $\{S_4|\frac{1}{2},\frac{1}{2},\frac{1}{2}\}$, $\{M_x|\frac{1}{2},\frac{1}{2},\frac{1}{2}\}$, $\{M_y|\frac{1}{2},\frac{1}{2},\frac{1}{2}\}$, $\{C_{2x}|\frac{1}{2},\frac{1}{2},\frac{1}{2}\}$, $\{C_{2y}|\frac{1}{2},\frac{1}{2},\frac{1}{2}\}$ connect the SSMPs.
In other words, there exist $\{\Theta C_4|\frac{1}{2},\frac{1}{2},\frac{1}{2}\}$, $\{\Theta S_4|\frac{1}{2},\frac{1}{2},\frac{1}{2}\}$, $\{\Theta M_x|\frac{1}{2},\frac{1}{2},\frac{1}{2}\}$, $\{\Theta M_y|\frac{1}{2},\frac{1}{2},\frac{1}{2}\}$, $\{\Theta C_{2x}|\frac{1}{2},\frac{1}{2},\frac{1}{2}\}$ and $\{\Theta C_{2y}|\frac{1}{2},\frac{1}{2},\frac{1}{2}\}$ symmetries. The existence of these symmetries ensure an alternating four-quadrant spin-split pattern in the Brillouin Zone \cite{yuan2020giant}, and protect the spin degeneracy along the nodal lines and nodal planes including the $\Gamma$ point \cite{PhysRevB.109.024404}.
Lifting of the SSMP interconverting symmetries in the rutile antiferromagnet can lead to the degeneracy lifting on the nodal lines and nodal planes, even at $\Gamma$.
The momentum-dependence of the spin splitting can be further attributed to the distinct area geometries (diamond versus rectangular geometry, when projected onto the (110) plane (\autoref{fig:fef2_vca_plus_sro}a) \cite{yuan2023degeneracy}. 
This geometric difference corresponds to an uncompensated magnetization within the (110) planes in real space and manifests as spin splitting of the energy bands along the [110] direction in $k$-space. 
The contrast in SSMP geometry thus provides an intuitive graphical framework for predicting momentum-dependent NRSS in collinear AFMs.

\subsection{NRSS in short range ordered FeOF}

\subsubsection{NRSS in FeOF VCA model}
FeOF crystallizes in the rutile-type structure like like FeF$_2$  with O$^{2-}$ and F$^-$ ions sharing the $4f$ anion sites \cite{brink2001nonstoichiometric,vlasse1973refinement}.
While FeOF does not exhibit long-range anion order, multiple experiments point to robust short-range correlations. 
Mössbauer spectroscopy has established that Fe consistently resides in \emph{fac} [FeO$_3$F$_3$] octahedra and orders antiferromagnetically  \cite{chappert1966effet,chappert1966structure}. 
Electron diffraction has further revealed extended anion correlations confined to the \{110\}, \emph{i.e.}, both (110) and (1$\overline{1}$0) planes \cite{brink2000electron}, accompanied by off-center displacements of Fe atoms, but with little correlation from plane to plane (illustrated in \autoref{fig:intro}b). 
These findings indicate that while the average structure retains tetragonal rutile symmetry, FeOF exhibits pronounced short-range ordering (SRO) of O$^{2-}$ and F$^-$ anions within each (110) plane that lead to local symmetry reductions.

A convenient way to model this chemical disorder is to replace the mixed anion site with a composition-averaged ``virtual'' atom.
This approach is known as the virtual crystal approximation (VCA) \cite{bellaiche2000electronic}.
Because the VCA uses an averaged pseudopotential for the F/O sites, FeOF retains the same symmetry as the fully ordered FeF$_2$ and is therefore expected to exhibit a similar momentum-dependent spin splitting along $\Gamma$-M.
This is confirmed by the calculated bands in \autoref{fig:fef2_vca_plus_sro}b.
However, the spin splitting in FeOF is significantly larger than in FeF$_2$, reaching a maximum of 432 meV for the bands at the Fermi level. We attribute this enhancement to stronger Fe-X (X=F/O) hybridization. 
Consistently, the calculated on-site magnetic moments on Fe$^{3+}$ are about $\approx 3.5 \mu_B$, indicating an occupation nominally higher than the ideal high spin $d^5$ state -- likely due to increased delocalization driven by the Fe-X hybridization, as evidenced in the projected DOS (\autoref{fig:dos_fef2_vca}b). 
The strong hybridization explains the enhanced superexchange through virtual hopping via Fe-X-Fe, which also significantly increases the N\'eel temperature \cite{chamberland1970preparation}.
Last, we note that because both the valence and conduction bands edges are derived from the Fe $3d$ orbitals, FeOF is classified as a Slater insulator \cite{PhysRev.82.538} rather than a charge transfer semiconductor 
\cite{chevrier2013first}.

\begin{figure*}
    \centering
    \includegraphics[width=0.97\textwidth]{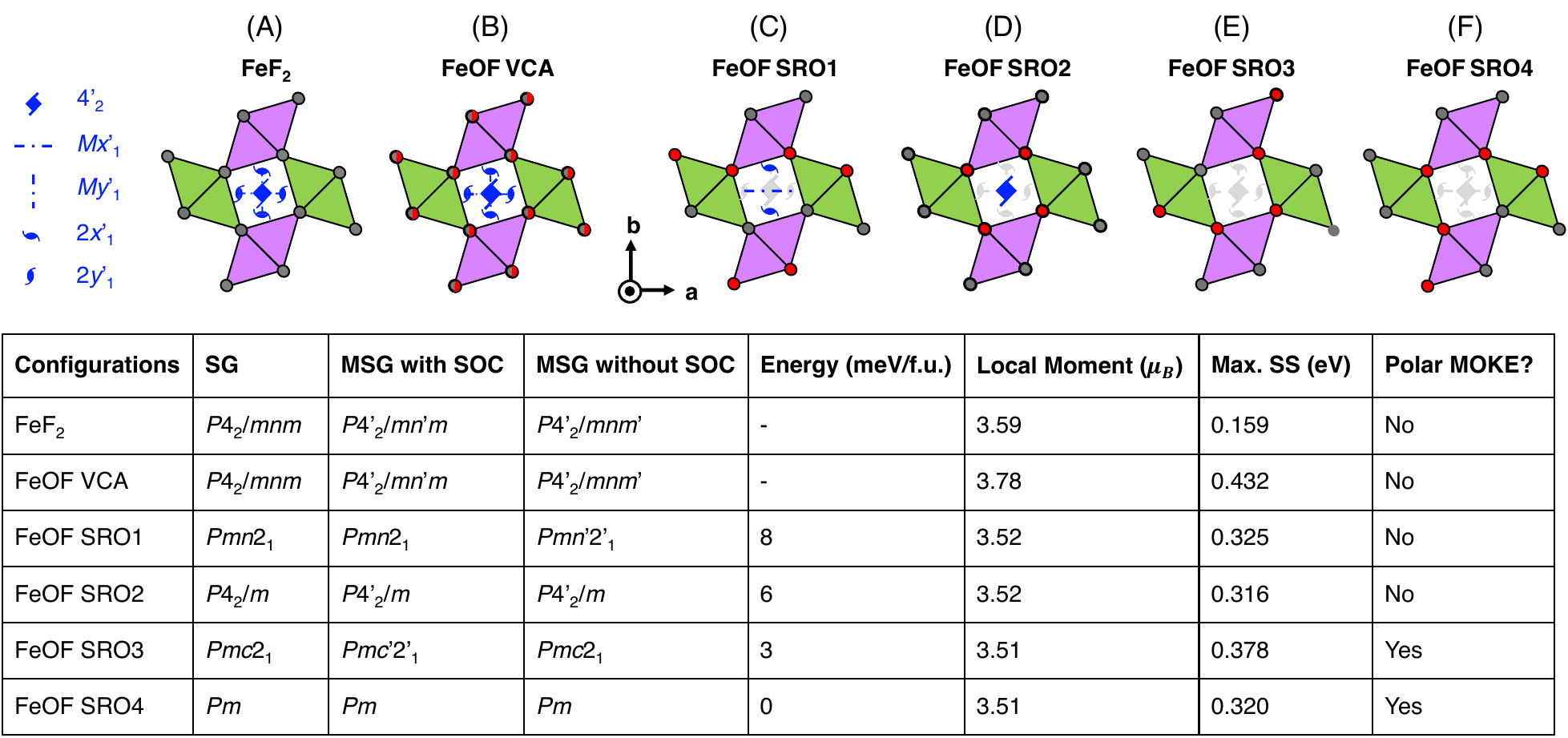}
    \caption{Schematic representation of rutile FeF$_2$ and FeOF models (top-down view) illustrating the main symmetries connecting the two opposite spin sublattices and structural motifs (A–F). Preserved symmetries are shown in blue, while broken symmetries are shown in gray.
    The symmetry operations $4_2^\prime$, $Mx^\prime_1$, $My^\prime_1$, $2x^\prime_1$, and $2y^\prime_1$ corresponds to $\{\Theta C_4|1/2,1/2,1/2\}$, $\{\Theta M_x|1/2,1/2,1/2\}$, $\{\Theta M_y|1/2,1/2,1/2\}$, $\{\Theta C_{2x}|1/2,1/2,1/2\}$, and $\{\Theta C_{2y}|1/2,1/2,1/2\}$, respectively.
    The lower panel summarizes the symmetry group, relative energy, spin splitting, and key properties of the considered configurations. Magnetic space groups (MSGs) with and without SOC are listed as the combined space–time symmetry groups of the magnets. The difference between the two cases arises from whether spatial and spin symmetry operations are coupled (with SOC) or decoupled (without SOC). The value of the maximum spin splitting (SS) is given for the top two valences bands along $\Gamma-\mathrm{M}$, and the value is averaged over multiple fractional weighted unfolded bands for SRO models.
    }
    \label{fig:ssmp}
\end{figure*}

\subsubsection{Atomistic description of SRO in FeOF}
While the VCA model reasonably reproduces the magnetic properties, it does not explicitly account for the symmetry breaking introduced by the anion SRO. 
To understand how anionic SRO affects the NRSS of the energy bands in FeOF, we used a CEM+MC simulation framework obtain equilibrium FeOF structures.
Our approach successfully reproduce the experimental O/F correlation on $\{110\}$ planes by annealing from 8000K to 0K using a canonical simulated annealing approach.
We find multiple competing low-energy structures belonging to four space group symmetries (\autoref{tab:table_mc}): $Pmn2_1$, $Pmc2_1$, $P4_{2}/m$, $Pm$.
Energetically, these four configurations are nearly identical (with energy differences of upto only $\approx$ 8 meV per formula unit,  \autoref{fig:ssmp}).
This indicates strong competition among them and is attributed to the lack of long-range order. 
These competing structures maintain the same ordering patterns along their $\{ 110 \}$ planes but exhibit subtle variations in the plane-to-plane (P2P) correlations of the O/F atoms. (A quantitative analysis of the difference is provided in \autoref{sec:mc}).

A closer examination of the four low-energy configurations confirms two shared features:
First, the presence of \emph{fac} [FeO$_3$F$_3$] octahedra, where Fe atoms displace towards the oxygen-rich face. 
Second, an alternating O/F atomic pattern along the [001] direction within each (110) plane: consecutive octahedra possess alternate faces consisting of a shared $\cdots$O-F/O-F$\cdots$ pattern along the rutile [001] direction as indicated in (\autoref{fig:intro}b).
Such anion correlation is consistent with previous simulations and experiments on FeOF \cite{brink2000electron,chevrier2013first}. Similar anion correlation was shown in TiOF \cite{PhysRevMaterials.8.054602,cumby2018high}, but arranged in different pattern.
Brink et al.\ attributed the emergence of the SRO to arise from bond-valence considerations \cite{brink2000electron}, but we pointed out, very recently, the anionic SRO in rutile oxyfluordes may have an electronic origin \cite{PhysRevMaterials.8.054602}. 
In FeOF, the nominal Fe$^{3+}$ cation exhibits a high spin d$^5$ state. Because this d$^5$ high spin has no partially filled/empty orbital pair that can stabilize an Fe–Fe dimers, FeOF forms an anion-ordering pattern (\emph{i.e.}, the shared $\cdots$O-F/O-F$\cdots$ arrangement) to remove the instability that would otherwise drive a Peierls-like distortion.
Since the four configurations sufficierntly capture the anionic SRO pattern within their $\{110\}$ planes and are computationally inexpensive, we use them for our subsequent  electronic structure studies.

\subsubsection{NRSS in FeOF SRO models}

\autoref{fig:fef2_vca_plus_sro} compares the calculated electronic structures of the four FeOF SRO models with those of FeF$_2$ and the FeOF VCA model. 
The energy bands are unfolded onto the same primitive FeF$_2$ cell to enable a direct comparison \cite{popescu2010effective,ku2010unfolding}. 
For the SRO models, we used G-AFM ordering, consistent with  FeF$_2$, as indicated by our DFT calculations  which show the G-AFM state is 0.5-1.5 eV/f.u.\ more stable than any  other collinear magnetic configuration. 
One one hand, similar to the long-range ordered rutile FeF$_2$ and the VCA model of FeOF, the band structures of all four configurations exhibit large spin splitting along the $\Gamma-\mathrm{M}$ direction, corresponding to to the [110] crystallographic direction. 
This splitting is a direct consequence of the distinct shapes (diamond vs.\ rectangle) of the SSMPs associated with the two opposite spin sublattices. 
Furthermore, the consistent O/F correlations across unit cells in the \{110\} planes reinforces the separation of the spin-polarized energy bands.
The magnitude of spin splitting remains fairly uniform  across the four configurations, with the maximum spin splitting at the Fermi level ranging from approximately 320 to 380 meV (\autoref{fig:ssmp}). 
This is greater than the spin splitting calculated in  FeF$_2$, which can be attributed to the stronger Fe--O hybridization in FeOF compared to FeF$_2$. 
However, these values are smaller than those obtained with the VCA model. This can be explained by comparing the density of states (DOS) of the VCA model with those of the SRO models. 
\autoref{fig:dos_fef2_vca}b clearly shows that the VCA treats the anion states using a pseudopotential averaged over O/F, resulting in an overestimation of Fe-F hybridization, as well as the the magnitude of spin splitting. Conversely, the DOS of FeF$_2$ (\autoref{fig:dos_fef2_vca}a) and the SRO1 $Pmn2_1$ configuration (\autoref{fig:dos_fef2_vca}c) show that the F states are localized deeper (-3\,eV) in the valence band and do not hybridize with the Fe states.

On the other hand, despite the subtle variations in the O/F correlations among the SRO models, their electronic structures exhibit notable differences. 
Spin splitting occurs on and along additional high symmetry $k$-points and paths that are spin degenerate in FeF$_2$ and the FeOF VCA model.
Specifically, the $P4_2/m$ configuration shows spin splitting along $\Gamma-\mathrm{X}$ and $\mathrm{X}-\mathrm{M}$ (\autoref{fig:fef2_vca_plus_sro}b), while spin splitting at the $\Gamma$ point emerges in the $Pmc2_1$ (\autoref{fig:fef2_vca_plus_sro}c) and $Pm$ (\autoref{fig:fef2_vca_plus_sro}d) configurations.

The momentum dependent spin splitting is a direct consequence of the symmetry breaking imposed by the different anionic SROs.
\autoref{fig:ssmp}a--f displays the crystal and magnetic symmetries that relate the spin-up and spin-down SSMPs.
Applying the same analysis as for FeF$_2$ (FeOF VCA exhibits the same symmetry), we find that in the $Pmn2_1$ configuration (\autoref{fig:fef2_vca_plus_sro}c, the spin-up (magenta) and spin-down (green) motifs (SSMPs) are connected by the operations \{$C_{2y}$ $\vert \frac{1}{2}, \frac{1}{2}, \frac{1}{2}\} $ and \{$M_x$ $\vert \frac{1}{2}, \frac{1}{2}, \frac{1}{2}\} $ (\autoref{fig:ssmp}c).
These operations leave all wave vectors lying on constant  $k_{x}$ planes, including $\Gamma-\mathrm{X}$, $\mathrm{X}-\mathrm{M}$, and $\mathrm{R}-\mathrm{A}$ (not shown) and $k_y$ parallel lines ($\mathrm{R}-\mathrm{A}$, not shown) invariant, resulting in spin degeneracy of the energy bands along the corresponding high-symmetry $k$ paths (\autoref{fig:fef2_vca_plus_sro}c).
In the $P4_2/m$ configuration, the spin-up and spin-down motifs are related by two pairs of operations: \{$C_{4z}^{+}$ $\vert \frac{1}{2}, \frac{1}{2}, \frac{1}{2}\} $ and \{$C_{4z}^{-}$ $\vert \frac{1}{2}, \frac{1}{2}, \frac{1}{2}\} $, which correspond to counterclockwise and clockwise four-fold rotations about [001], followed by a translation \{$\frac{1}{2}, \frac{1}{2}, \frac{1}{2}\}$. The second pair, \{$S_{4z}^{+}$ $\vert \frac{1}{2}, \frac{1}{2}, \frac{1}{2}\} $ and \{$S_{4z}^{-}$ $\vert \frac{1}{2}, \frac{1}{2}, \frac{1}{2}\}$, are similar and include an additional reflection about the [001] $z$ axis, following counterclockwise and clockwise four-fold rotations,  respectively (\autoref{fig:ssmp}d). 
By contrast, the specific O/F ordering in the $Pmc2_1$ and $Pm$ SRO configurations eliminate all rotational and mirror symmetries that connect the spin-up and spin-down octahedra motifs. This absence of such spin-linking symmetries leads to spin splitting across all high-symmetry $k$-paths and even at the $\Gamma$ point (\autoref{fig:fef2_vca_plus_sro}e,f). 
Materials exhibiting $\Gamma$ spin splitting behave more like ferromagnets or ferrimagnets and may enable the generation of spin currents without cancellation arising from the alternating spin polarizations \cite{yuan2024nonrelativistic}.

\section{Discussion}

\subsection{Detecting the SRO-induced NRSS effect}

 \begin{figure}
    \centering
    \includegraphics[width=0.49\textwidth]{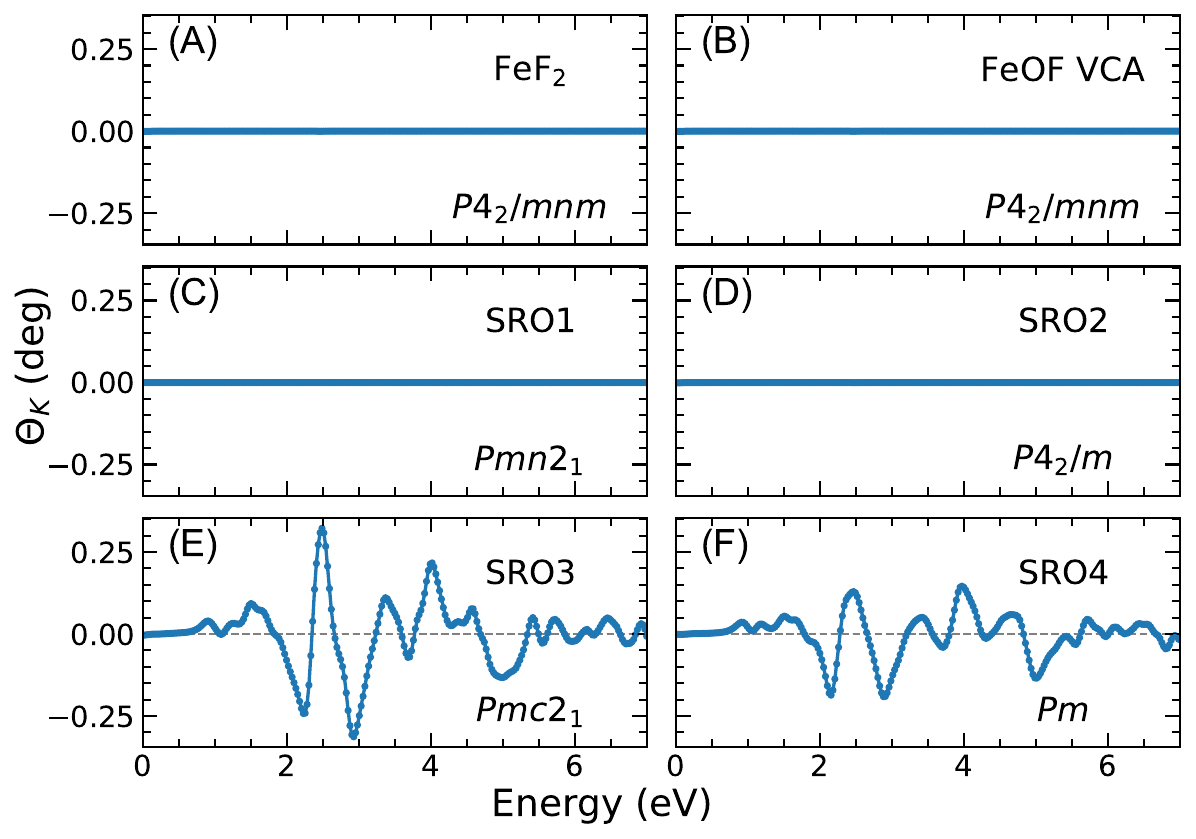}
    \caption{Calculated Kerr rotation spectra for all structures including the null spectra (a) FeF$_2$, (b) VCA model of FeOF and two of our selected configurations; (c) $Pmn2_1$ and (d) $P4_2/m$. (e) and (f) show the nonzero MOKE responses of the $Pmc2_1$ and $Pm$ configurations, respectively.}
    \label{fig:moke}
\end{figure}

Since the DFT and DFT$+U$ calculations give a band gap ranging from 1\,eV ($U = 0$\,eV) to $\sim$ 2\,eV ($U = 4$\,eV) (\autoref{sec:u}, \autoref{fig:dos_bands_u}), we believe the NRSS effect in FeOF should be optically accessible, particularly through the magneto-optical Kerr effect (MOKE). Using the frequency-dependent dielectric tensor obtained from our first-principles calculations, we evaluated the polar MOKE response under normal incidence. Among the four FeOF SRO configurations considered, only the $Pm$ and $Pmc2_1$ structures -- both  exhibiting spin splitting at $\Gamma$ -- show finite off-diagonal dielectric components, consistent with their magnetic space groups with SOC ($Pm$ and $Pmc'2_1'$) \cite{cheong2023trompe}. These components produce nonzero Kerr rotations across a broad spectral range (\autoref{fig:moke}e,f). By contrast, the $\Gamma$-degenerate configurations with symmetries $Pmn'2_1'$ and $P4_2'/m$ (\autoref{fig:moke}c,d), as well as FeF$_2$ and the VCA FeOF model ($P4_2'/mnm'$), exhibit no Kerr signal (\autoref{fig:moke}a,b), exactly as symmetry forbids \cite{cheong2023trompe,watanabe2024symmetry}. The variation in MOKE responses can be understood by examining  how the antisymmetric component of the calculated frequency-dependent dielectric tensor transforms under symmetry operations. For example, in FeF$_2$ with magnetic group  $P4_2'/mnm'$, the mirror symmetry $M_x$ and/or $M_y$ enforce a zero Kerr rotation about z. 

In a realistic FeOF specimen, anionic SRO is expected to produce a heterogeneous morphology composed of nanoscale regions locally resembling the motifs examined here. We anticipate that the  measured Kerr spectrum would approximate a volume-fraction–weighted average of the local MOKE responses. Given that $\Gamma$-split NRSS AFMs possess magnetic symmetries that can also describe ferromagnetic-like behavior as a consequence of the broken $UR$ symmetry, MOKE is a promising experimental probe to examine the connection between SRO and its impact on the local electronic structure in FeOF single crystals. 
As described in \autoref{sec:mc} and illustrated in \autoref{fig:grid_emd}, our Monte Carlo simulations indicate that the space group of the annealed structure reduces to $P1$ at a supercell size of $7 \times 7 \times 7$, corresponding to an approximate SRO domain length of 2–3 nm. We therefore propose using extreme anti-reflection enhanced MOKE microscopy, which has demonstrated the ability to reslove magnetic domains in 1 nm-thick Co films \cite{kim2020extreme}, to probe the influence of SRO on the nanoscale spin dynamics in FeOF.

\subsection{Oxyfluorides as high N\'eel temperature NRSS AFMs}
By considering only the first exchange interaction and fitting a Heisenberg model to the energies of two magnetic orientations in the $Pmn2_1$ configuration of FeOF, we estimate the first magnetic exchange constant $J_1 = 3.24$\,meV. Using the molecular mean-field approximation \cite{anderson1950generalizations}, this yields a N\'eel temperature of approximately 438 K for FeOF---within an order of magnitude of the experimentally measured 315 K and significantly higher than that of its single-anion rutile counterpart, FeF$_2$. We attribute this enhancement to the Fe-O-Fe linkage enabling indirect virtual hopping through both $p\sigma$ and $p\pi$ orbitals, resulting in stronger superexchange compared to FeF$_2$ \cite{chamberland1970preparation}.

\begin{figure}
    \centering
    \includegraphics[width=0.49\textwidth]{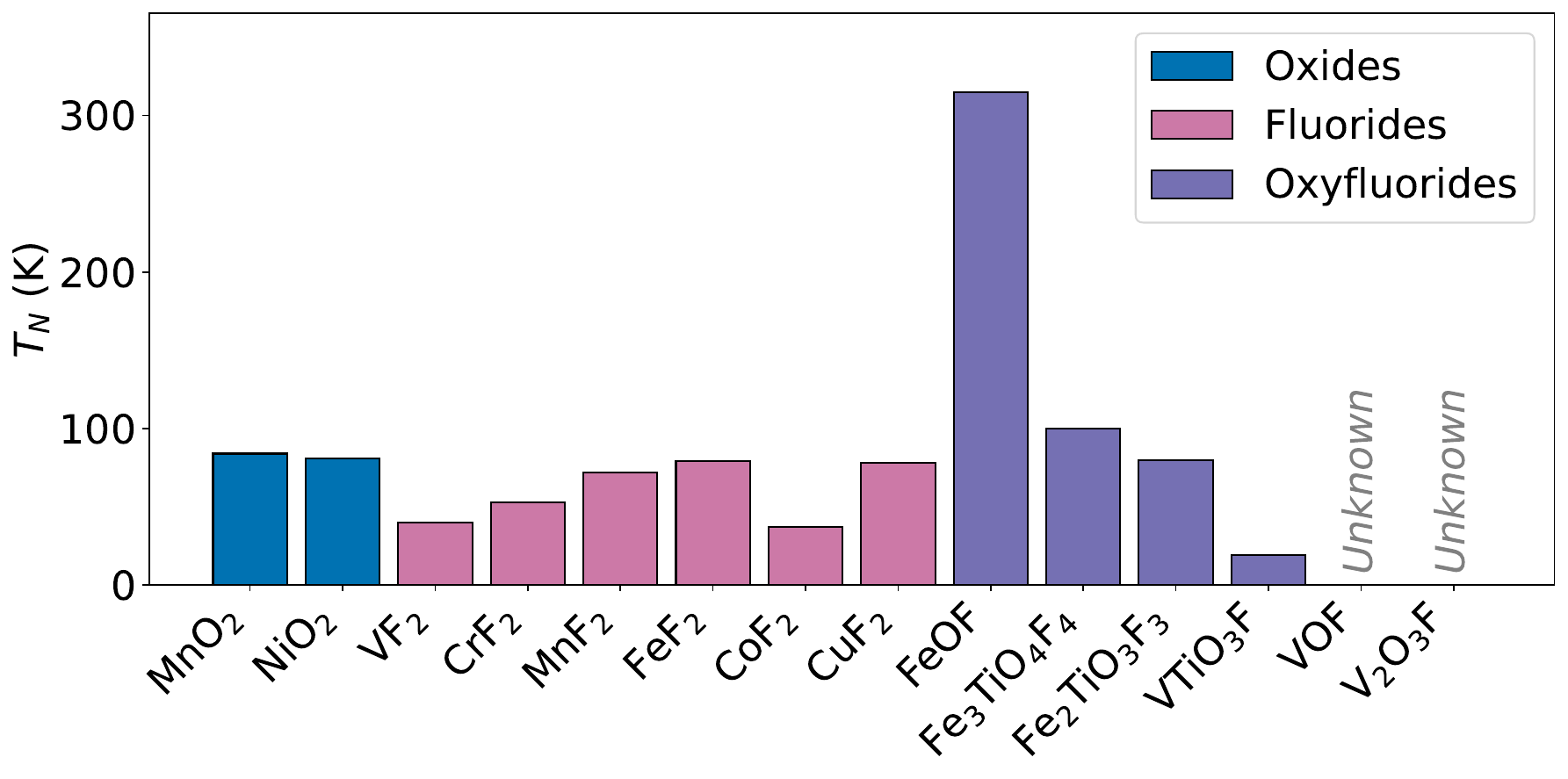}
    \caption{N\'eel temperatures of antiferromagnetic rutile oxides, fluorides, and oxyfluorides. Unlike oxides and fluorides, magnetic characterization data for many rutile oxyfluorides is incomplete. All values taken from Ref.~\cite{chamberland1970preparation}.}
    \label{fig:neel_temps}
\end{figure}

As shown in \autoref{fig:neel_temps}, FeOF exhibits a N\'eel temperature substantially higher than other antiferromagnetic rutile compounds, most of which are fluorides or oxides. The scarcity of data on magnetic ordering in other rutile oxyfluorides highlights the need for a systematic investigation of this class of materials. Heteroanionic compounds, by enabling favorable transition-metal cation physics beyond what is achievable in homoanionic systems, offer a promising route to designing room-temperature NRSS materials. The lack of experimental synthesis and magnetic characterization for many candidate antiferromagnetic oxyfluorides (\autoref{fig:neel_temps}) further underscores the importance of comprehensive studies on this family.

\section{Conclusion}
We investigated the impact of short-range anionic order (SRO) on the electronic structure of FeOF using four representative configurations. Our results reveal robust momentum-dependent spin splitting along the $\Gamma-\mathrm{M}$ direction, closely resembling the nonrelativistic spin-splitting (NRSS) fingerprints of long-range-ordered FeF$_2$. Although all of our chosen configurations exhibit a similar magnitude of spin splitting, the character of the splitting—whether it occurs at $\Gamma$ or not—depend sensitively on directional O/F correlations, effects absent in  FeF$_2$ and not captured by the virtual crystal approximation (VCA) model of FeOF.

Because these configurations are energetically similar, real FeOF specimens are likely to exhibit nanoscale domains with plane-to-plane O/F correlations rather than a long-range ordered superstructure. This expectation is supported by our large-supercell annealing simulations (\autoref{tab:table_mc}), which yield $P1$ symmetry 
with no medium or long-range order. 
Consequently, experimentally synthesized FeOF should display the robust $\Gamma-\mathrm{M}$ spin splitting common to all configurations, along with additional symmetry-dependent splittings at other high-symmetry $k$ points, including $\Gamma$ in certain motifs.

We also discussed experimental strategies to probe SRO and its influence on NRSS behavior. In particular, we propose extreme anti-reflection-enhanced MOKE microscopy as a promising technique to resolve nanoscale domains and correlate local band structure with SRO. Finally, we highlighted that antiferromagnetic heteroanionic materials remain an underexplored but promising class of NRSS compounds. Theoretical insights presented here provide a foundation for experimental verification and suggest that anion substitution can serve as a powerful design parameter for tuning spin-related responses in functional materials.

\begin{acknowledgments}
S.S.N.\ and J.M.R.\ were supported by the National Science Foundation (NSF) under award number DMR-2413680. D.P.\ was supported by the SUPeRior Energy-efficient Materials and dEvices (SUPREME) Center SUPREME, one of seven centers in the JUMP 2.0, a Semiconductor Research Corporation (SRC) program sponsored by DARPA consortium. L.Y. was supported by the Air Force Office of Scientific Research under (AFOSR) Award No.\ FA9550-23-1-0042. The authors are grateful to Prof.\ James Cumby for discussions on anionic SRO in FeOF. DFT calculations were performed using the \texttt{Quest} HPC facility at Northwestern and \texttt{Bridges-2} at the Pittsburgh Supercomputing Center from the Advanced Cyberinfrastructure Coordination Ecosystem: Services and Support (ACCESS) program, which is supported by NSF grants 2138259, 2138286, 2138307, 2137603, and 2138296.
\end{acknowledgments}

\section*{Appendix}
\appendix
\renewcommand{\thefigure}{A\arabic{figure}} %
\renewcommand{\thetable}{A\Roman{table}} 

\setcounter{figure}{0} 
\setcounter{table}{0} 

\begin{figure*}
    \centering
    \includegraphics[width=0.65\textwidth]{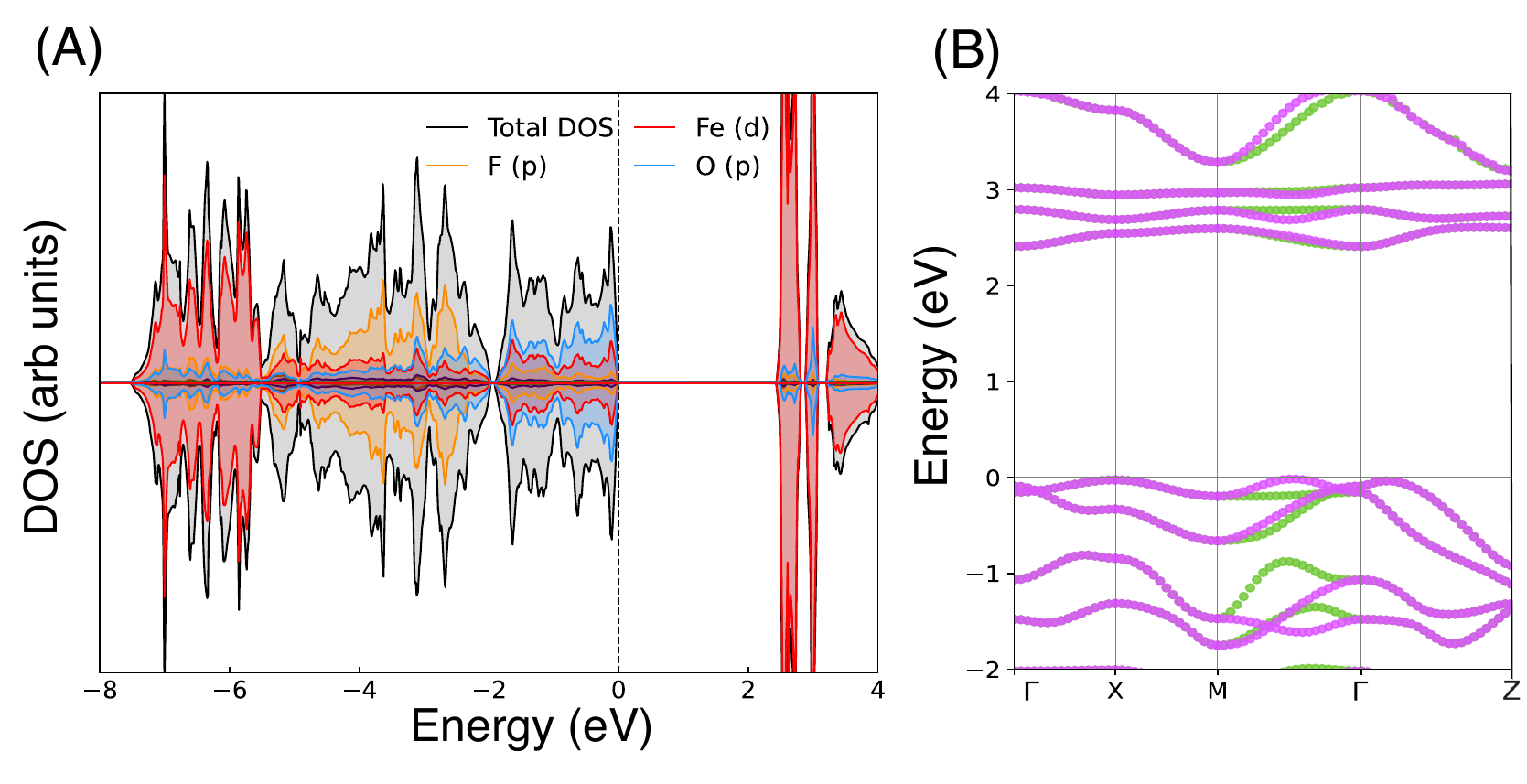}
    \caption{(a) Calculated density of states (DOS) for the FeOF SRO1 ($Pmn2_1$) structure using the GGA$+U$ approach with $U_{\text{eff}} = 4$\,eV. Electron-electron interactions shift the Fe $3d$ states deeper into the valence band, while the oxygen $2p$ states dominate near the valence-band maximum. This enhanced $p$ character reduces carrier mobility near the Fermi level and diminishes spin splitting. (b) Unfolded band structure projected onto the primitive rutile FeF$_2$ cell, where pink and green curves denote spin-up and spin-down states, respectively.}
    \label{fig:dos_bands_u}
\end{figure*}

\subsection{Symmetry conditions for NRSS}\label{sec:symm}
We briefly review the symmetry requirements for NRSS in collinear AFMs.
NRSS in collinear AFM materials is enabled by the breaking of both $\Theta I$ and $UT$ symmetries \cite{yuan2021prediction,yuan2020giant}.
Here, the symmetry operation $U$ refers to a spin rotation of the SU(2) group acting on the spin 1/2 space, which simply reverses the (collinear) spin; $T$ is a spatial translation; $\Theta$ refers to the time-reversal operation and $I$ represents spatial inversion \footnote{The symmetry analysis is consistent with the spin symmetry framework used in Refs.\ \cite{PhysRevX.12.021016,vsmejkal2020crystal}. These two symmetries maps to ${C_2||E|t}$ and ${-1||-1}$ following Litvin's notation \cite{litvin1977spin}.}
Classifying materials based on how these two conditions are satisfied or violated, leads to various AFM prototypes with spin-degenerate or spin-split bands. 
AFMs breaking both symmetry conditions exhibit NRSS and form a distinct category (SST-4) \cite{yuan2021prediction,yuan2020giant}, separate from ferromagnets and spin degenerate conventional AFMs.
The symmetry conditions can be understood in terms of the local geometric spin-structure motif pairs (SSMP), such as octahedra, each carrying opposite magnetic moments \cite{yuan2023degeneracy}. 
When the constituent SSMP are inter-convertible, meaning they are related by the inversion or translation (this is equivalent to preserving $\Theta I$ or $UT$ symmetry \cite{yuan2020giant,yuan2021prediction}), the AFM will exhibit degenerate energy bands. 
In contrast, AFMs with non-inter-convertible constituent SSMP, where there is no $\Theta I$ or $UT$ symmetry between the constituent SSMP, will exhibit NRSS. 
SST-4 materials can be divided further into subclasses depending on whether the SSMPs are connected to each other via auxiliary symmetries, \emph{e.g.}, rotation $R$ and/or mirror $m$, or not \cite{mrzv-wmcf}.
Materials with an auxiliary rotational symmetry are collinear altermagnets. The auxiliary symmetry enforces the spin-splitting to alternate in sign and along trajectories connected by $R$ in $k$-space, intersecting at the origin $\Gamma$. 
In contrast, collinear AFMs whose SSMPs are \emph{not} connected by the $UR$ operation (considering spin reversal also) will exhibit energy bands that are spin-split at the $\Gamma$ point \cite{yuan2024nonrelativistic}. 
This distinct subclass of NRSS materials are envisioned to yield non-trivial spin-based responses compared to altermagnetic materials.

\subsection{Effect of spin-orbit coupling}
The effect of spin-orbit coupling (SOC) on the degeneracy of the bands can be understood by examining the symmetry change upon including SOC in the symmetry analysis. 
We can define the magnetic space group (MSG) with (without SOC) as the  group of space and time symmetries of the compound when SOC is considered (neglected). They differ by whether the symmetry operations that act on the spatial space and the spin space are coupled (decoupled). 
The MSG with SOC is the same to the commonly defined MSG, which dictates the band degeneracy for materials considering SOC. The MSG without SOC is a nontrivial subgroup of the spin space group, which can be used describe the band degeneracy when SOC is neglected (\emph{i.e.}, nonrelativisitc limit).
For collinear magnets, the MSG without SOC is isomorphic to the nontrivial part of the spin space group. The analysis can be done using \texttt{FindMagSym} tool \cite{yuan2024nonrelativistic}. 
For the FeOF SRO2 ($P4_2/m$) and SRO4 ($Pm$) configurations, the  magnetic space group remains unchanged upon including SOC (\autoref{fig:ssmp}). For SRO1 ($Pmn2_1$), the MSG changes from type III to type I, while for SRO3 ($Pmc2_1$), the MSG changes from type I to type III. However, given that FeOF comprises of a $3d$ transition metal, we do not expect SOC to play a major role in influencing spin splitting.

\subsection{Effect of electron correlation DFT$+U$}\label{sec:u}
To evaluate the impact of electronic correlations on the FeOF band structure, we performed full atomic relaxations for all four candidate structures using the DFT$+U$ method. The Hubbard correction was implemented following the Dudarev approach \cite{dudarev1998electron}, with an effective parameter $U_{\text{eff}} = U - J = 4$\,eV applied to Fe$^{3+}$ sites, consistent with prior first-principles studies \cite{chevrier2013first}.
Using the $Pmn2_1$ configuration as a representative case, we find that introducing on-site electron–electron interactions shifts the Fe $3d$ states deeper into the valence band (\autoref{fig:dos_bands_u}a). Consequently, the valence-band edge becomes predominately oxygen $2p$ character, characteristic of a $p$–$d$ charge-transfer insulator, in agreement with \cite{chevrier2013first}. The bands near the Fermi level become significantly less dispersive, indicating larger effective masses, and exhibit reduced spin splitting. This flattening is evident in \autoref{fig:dos_bands_u}b.
Quantitatively, the maximum spin-splitting along $\Gamma - \mathrm{M}$ for the band crossing the Fermi level decreases from 0.325\,eV (without correlation) to 0.173\,eV with DFT$+U$. The calculated on-site magnetic moment on Fe atoms is 4.2\,$\mu_B$.

\begin{table}[t]
\centering
\caption{Calculated crystallographic parameters for FeOF SRO1 $Pmn2_1$ using the PBESol functional. Lattice parameters: $a = 2.9767$\,\AA, $b = 4.6054$\,\AA, $c = 4.6193$\,\AA.}
\begin{ruledtabular}
\begin{tabular}{ccccc}
 Atom & Site & $x$ & $y$ & $z$ \\
 \hline
 Fe & $2a$ & 0.00000 & 0.79513 & 0.00136 \\
 O  & $2a$ & 0.00000 & 0.05215 & 0.29808 \\
 F  & $2a$ & 0.00000 & 0.45197 & 0.70055 \\
\end{tabular}
\end{ruledtabular}
\end{table}

\begin{table}[t]
\centering
\caption{Calculated crystallographic parameters for FeOF SRO2 $P4_2/m$  using the PBESol functional. Lattice parameters: $a = b = 6.5075$\,\AA, $c = 2.9653$\,\AA.}
\begin{ruledtabular}
\begin{tabular}{ccccc}
 Atom & Site & $x$ & $y$ & $z$ \\
 \hline
 Fe & $4j$ & 0.73087 & 0.27088 & 0.00000 \\
 O  & $4j$ & 0.45251 & 0.24446 & 0.00000 \\
 F  & $4j$ & 0.04952 & 0.24157 & 0.00000 \\
\end{tabular}
\end{ruledtabular}
\end{table}

\begin{table}[t]
\centering
\caption{Calculated crystallographic parameters for FeOF SRO3 $Pmc2_1$ using the PBESol functional. Lattice parameters: $a = 2.9709$\,\AA, $b = 6.5115$\,\AA, $c = 6.5196$\,\AA.}
\begin{ruledtabular}
\begin{tabular}{ccccc}
 Atom & Site & $x$ & $y$ & $z$ \\
 \hline
 Fe & $2a$ & 0.00000 & 0.72912 & 0.02293 \\
 Fe & $2b$ & 0.50000 & 0.72977 & 0.47881 \\
 O  & $2a$ & 0.00000 & 0.75519 & 0.29946 \\
 O  & $2b$ & 0.50000 & 0.45138 & 0.49916 \\
 F  & $2a$ & 0.00000 & 0.75722 & 0.70162 \\
 F  & $2b$ & 0.50000 & 0.04933 & 0.49803 \\
\end{tabular}
\end{ruledtabular}
\end{table}

\begin{table}
\centering
\caption{Calculated crystallographic parameters for FeOF SRO4 $Pm$  using the PBESol functional. Lattice parameters: $a = 6.5134$\,\AA, $b = 2.9726$\,\AA, $c = 10.3031$\,\AA, with $\beta = 108.38^\circ$.}
\begin{ruledtabular}
\begin{tabular}{ccccc}
 Atom & Site & $x$ & $y$ & $z$ \\
 \hline
 Fe & $1a$ &  0.01414 & 0.00000 & -0.01424 \\
 Fe & $1b$ &  0.48646 & 0.50000 &  0.01551 \\
 Fe & $1a$ &  0.64043 & 0.00000 &  0.31928 \\
 Fe & $1b$ &  0.19576 & 0.50000 &  0.34842 \\
 Fe & $1a$ &  0.34701 & 0.00000 &  0.65211 \\
 Fe & $1b$ &  0.81888 & 0.50000 &  0.68196 \\
 O  & $1a$ &  0.29939 & 0.00000 &  0.00155 \\
 O  & $1b$ &  0.60661 & 0.50000 &  0.19970 \\
 O  & $1a$ &  0.36880 & 0.00000 &  0.33291 \\
 O  & $1b$ &  0.26005 & 0.50000 &  0.53279 \\
 O  & $1a$ &  0.63074 & 0.00000 &  0.66330 \\
 O  & $1b$ & -0.06922 & 0.50000 &  0.86671 \\
 F  & $1a$ &  0.69846 & 0.00000 & -0.00357 \\
 F  & $1b$ &  0.39875 & 0.50000 &  0.80097 \\
 F  & $1a$ & -0.03385 & 0.00000 &  0.33189 \\
 F  & $1b$ &  0.05928 & 0.50000 &  0.13460 \\
 F  & $1a$ &  0.03532 & 0.00000 &  0.66853 \\
 F  & $1b$ &  0.74300 & 0.50000 &  0.46759 \\
\end{tabular}
\end{ruledtabular}
\end{table}

\begin{figure}
    \centering
    \includegraphics[width=0.48\textwidth]{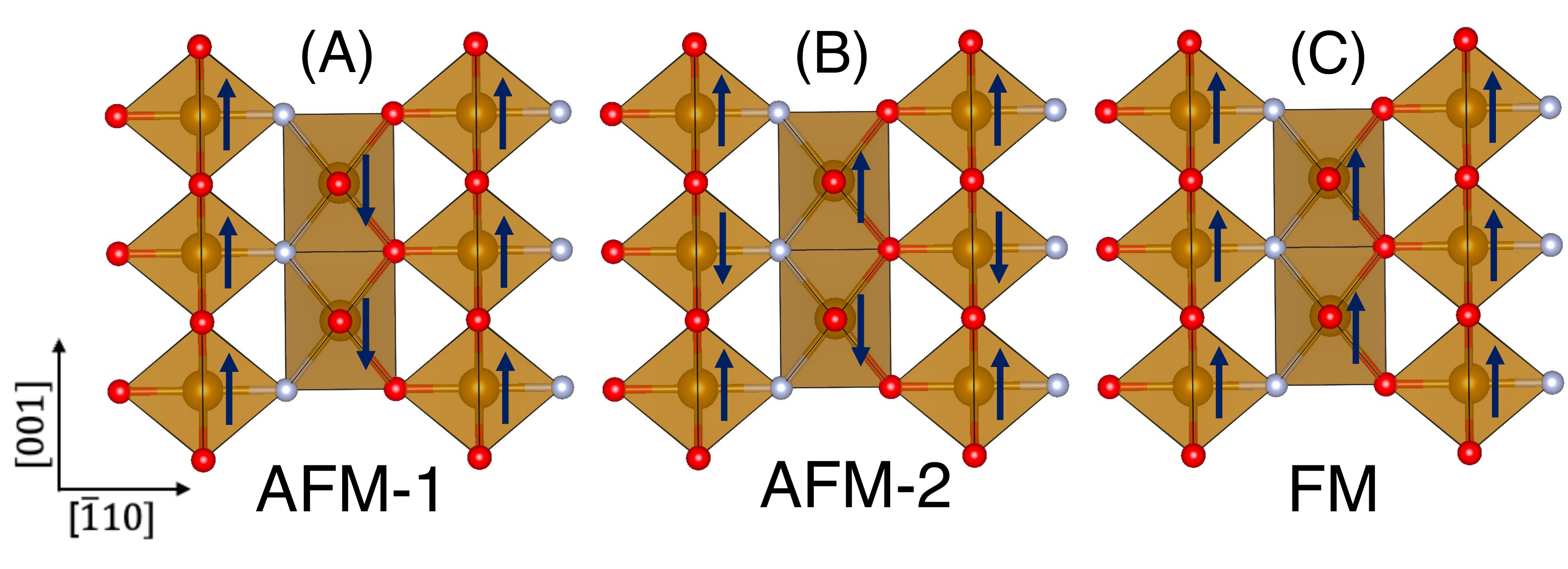}
    \caption{Simulated magnetic configurations for FeF$_2$ and all candidate FeOF structures, with dark blue arrows indicating spin orientation in a collinear arrangement. The AFM-1 configuration (a), corresponding to G-type antiferromagnetic order, is the most stable, consistent with the Goodenough-Kanamori superexchange rules. AFM-2 (b) is slightly less favorable, while the ferromagnetic configuration (c) is the least stable, as expected for a high-spin $d^5$ system such as FeOF.}
    \label{fig:mag_order}
\end{figure}

\begin{figure*}
    \centering
    \includegraphics[width=0.98\textwidth]{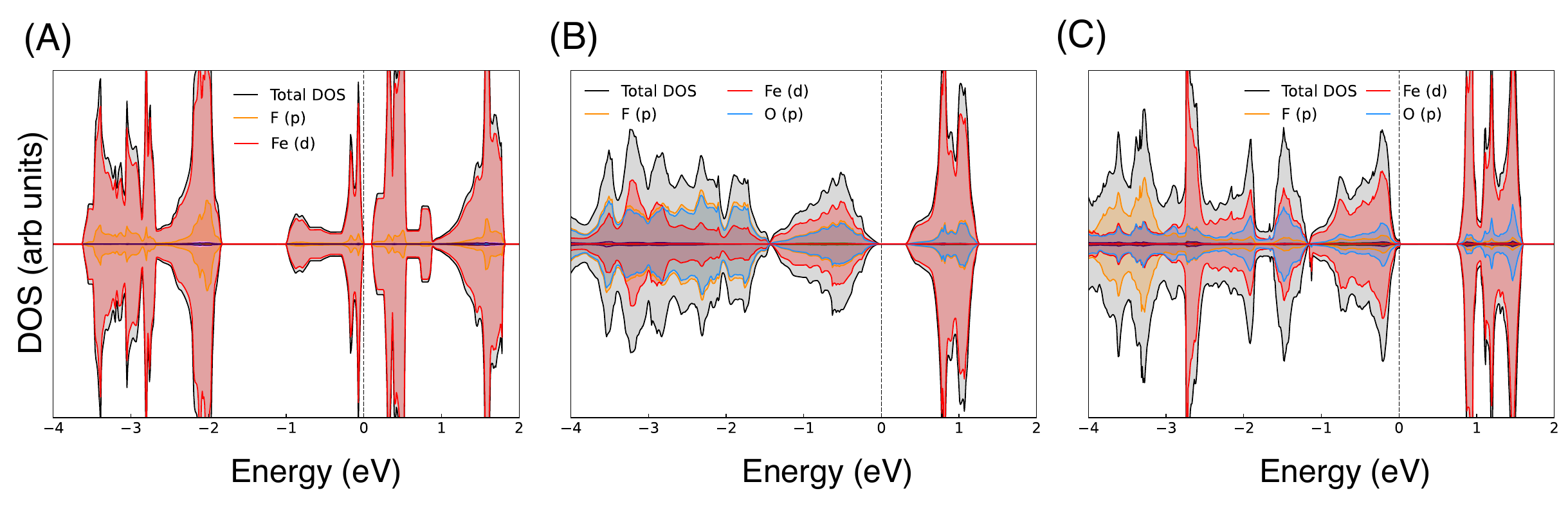}
    \caption{DFT-PBEsol density of states (DOS) for (a) FeF$_2$, (b) VCA model of FeOF, and (c) the SRO1 $Pmn2_1$ structure, illustrating that both FeF$_2$ and FeOF exhibit Slater-type insulating behavior. Unlike FeF$_2$, FeOF shows significantly stronger Fe-O hybridization than seen in the Fe-F interactions near the valence-band edge. Importantly, we see that the VCA model overestimates the extent of Fe-F hybridization in (b), leading to a spuriously high magnitude of spin splitting. The correct description of Fe-O hybridization is exhibited in (c), with the F states localized deeper in the valence band. }
    \label{fig:dos_fef2_vca}
\end{figure*}

\subsection{Modeling anion order in FeOF}
\label{sec:mc}
Heteroleptic coordination in oxyhalides is rarely random; short-range order (SRO) often emerges due to bonding preferences (\autoref{fig:intro}b), particularly along specific crystallographic directions such as [110]. These tendencies can be captured in atomic pair distribution functions (PDFs, \autoref{fig:intro}d–f) and are largely governed by transition-metal $d$-orbital filling. While PDFs reveal chemical ordering through peak shifts or modulations associated with short- and long-range correlations, the resulting differences in electronic structure are even more pronounced. Accurately modeling and controlling these local effects is therefore essential for exploiting NRSS in functional materials.

\autoref{tab:table_mc} summarizes the results of simulated annealing from 8000\,K to 300\,K for various FeOF supercell sizes. The ground states exhibit multiple space group symmetries ($Pmn2_1$,$Pmc2_1$, $P4_{2}/m$, $Pm$) for supercells ranging from $2 \times 2 \times 2$ (48 atoms) to $7 \times 7 \times 7$ (2058 atoms).
Beyond this size, the relaxed structures consistently adopt $P1$ symmetry, even for supercells as large as $40 \times 40 \times 40$. 
The presence of several competing low-energy structures identified by MC simulations prompted us to examine the DFT relaxations from the training set to understand the origin of this strong structural competition.

From the training data, we selected the smallest supercells corresponding to the symmetries $Pmn2_1$, $Pmc2_1$, $P4_{2}/m$, $Pm$ and compared their DFT-relaxed structures (see Tables AI–AIV). 
Analysis of the O-F ordering pattern reveals the expected [FeO$_3$F$_3$] \emph{fac} ordered octahedra previously reported for TiOF \cite{PhysRevMaterials.8.054602,cumby2018high}, MoON \cite{PhysRevLett.123.236402}, and observed experimentally and computationally for FeOF  \cite{brink2000electron,chevrier2013first}. 
In this arrangement, each alternate face contains exclusively O or F anions, enabling  Fe atoms to displace towards the O-rich face, as found by the shorter (longer) Fe--O (Fe--F) bonds. Furthermore, along each (110) plane in all  structures, consecutive octahedra  exhibit alternate faces consisting of a shared $\cdots$O-F/O-F$\cdots$ pattern along the rutile [001] direction. This is also consistent with previous structural characterizations \cite{brink2000electron} and previously rationalized using molecular orbital arguments \cite{PhysRevMaterials.8.054602}. 

Energetically, all four configurations are nearly degenerate, with differences of only $\approx$ 8 meV per formula unit, indicating strong competition among these states. Each configuration maintains the expected and consistent local ordering patterns along their respective (110) planes. The fundamental basic building units and local octahedral environments remain preserved throughout all structures. The main variations arise primarily in the plane-to-plane (P2P) O/F correlations.

\begin{table}
    \centering
    \begin{ruledtabular}
        \caption{Space groups of lowest-energy configurations for MC simulated annealing of different FeOF supercell sizes. Supercells greater than $7 \times 7 \times 7$ also resulted in $P1$ symmetry.}
    \begin{tabular}{c c}
        Supercell size &  Space group \\
        \hline\\[-0.8em]
        $2 \times 2 \times 2$ & $Pmc2_1$ \\
        $3 \times 3 \times 3$ & $P4_2/m$ \\
        $4 \times 4 \times 4$ & $Pmn2_1$ \\
        $5 \times 5 \times 5$ & $Pm$ \\
        $6 \times 6 \times 6$ & $Pm$ \\
        $7 \times 7 \times 7$ & $P1$ \\
        \end{tabular}
    \end{ruledtabular}
    \label{tab:table_mc}
\end{table}

\begin{figure*}
    \centering
    \includegraphics[width=0.75\textwidth]{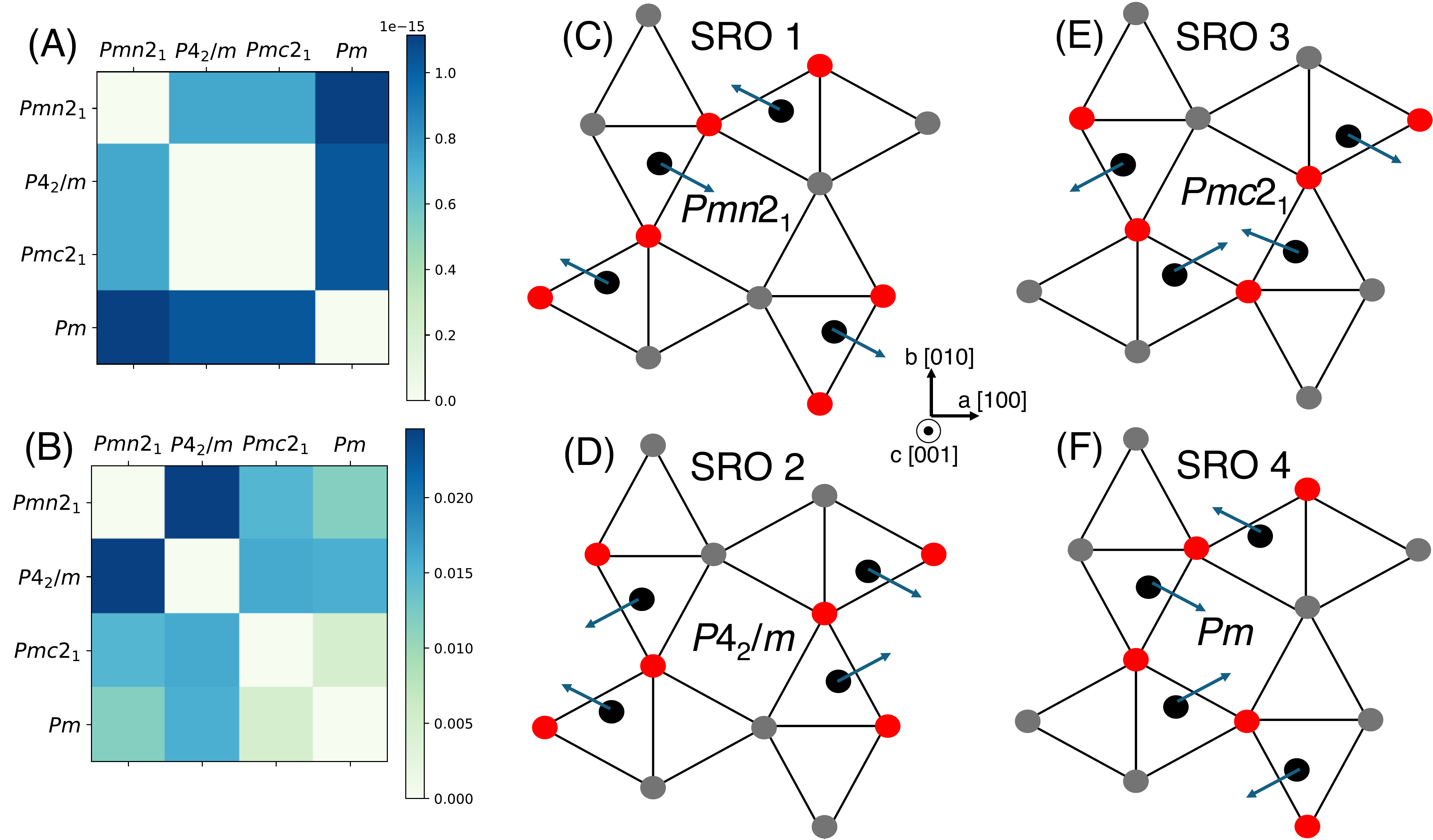}
    \caption{Pairwise comparison of Earth Mover’s Distance (EMD) values computed from GRID for all four candidate structures (a) before and (b) after structural relaxation. The increase in EMD following relaxation reflects the influence of O/F correlations on Fe displacements, as shown in (c–f), where Fe atoms (black) shift toward O atoms (red) and away from F atoms (gray).}
    \label{fig:grid_emd}
\end{figure*}

To capture these subtle P2P variations, we employed the Grouped Representation of Interatomic Distances (GRID) descriptor, developed by Zhang et al.\ \cite{zhang2023grouped}.
GRID extends the concept of a one-dimensional radial distribution function (RDF) by encoding additional structural information beyond simple pairwise distances, \emph{e.g.}, coordination environments, making it well-suited for distinguishing variations in P2P correlations that manifest as distinct symmetry elements.
The procedure for computing GRID is as follows:
\begin{enumerate}
    \item For each atom in the unit cell, compute all pairwise distances up to a chosen cutoff (10~\AA\ in this work), accounting for periodic boundary conditions.
    \item Rank these distances in ascending order to form a list for each atom:
    \[
        d_j = (d_{1j}, d_{2j}, \dots, d_{ij}, \dots)
    \]
    where the $i$-th distance is assigned to the $i$-th GRID group.
    \item Discretize the distances into binned histograms for each group and apply Gaussian kernel density estimation to smooth the distribution, avoiding discontinuities between bins.
\end{enumerate}

The Earth Mover's Distance (EMD) provides a robust quantitative metric for assessing structural (dis)similarities among FeOF short-range ordered (SRO) configurations. Conceptually, EMD represents the minimum amount of work required to transform one distribution into another—in this case, converting the structural distribution of one configuration (defined by Wyckoff site arrangements) into that of another. This approach enables a detailed comparison of O/F ordering patterns by quantifying the ``cost'' of switching between different coordination environments, particularly across extended scales such as the (110) planes.

For each pair of structures, EMD values were computed between corresponding ($i_{\text{th}}$–$i_{\text{th}}$) GRID distributions across the first 100 GRID groups. The mean EMD was then obtained by averaging over these groups, following the procedure outlined in \cite{zhang2023grouped}. This analysis converts subtle variations in O/F positioning and coordination environments into a measurable quantity, offering clear insights into how anion arrangements influence plane-to-plane correlations in these closely related structures.

\autoref{fig:grid_emd}a illustrates the GRID-based dissimilarity among four structures derived from the idealized rutile FeF$_2$ configuration prior to relaxation. These enumerated structures exhibit high similarity, with nearly indistinguishable GRID representations. Nevertheless, subtle differences in P2P O/F correlations lead to distinct displacement directions of Fe atoms from their idealized positions, although all structures consistently relax toward oxygen-containing faces (\autoref{fig:grid_emd}c–e).

After structural relaxation, \autoref{fig:grid_emd}b reveals Earth Mover's Distance (EMD) values in the range of $\approx$0.01-0.02 between any two structures. The increased dissimilarity post-relaxation arises from structural differentiation driven by O/F correlation differences present in the unrelaxed configurations. This progression, from nearly identical initial GRID representations (\autoref{fig:grid_emd}a) to distinct yet largely similar relaxed configurations (\autoref{fig:grid_emd}b), demonstrates how minor initial variations become more pronounced through Fe atom relaxation toward oxygen-rich octahedral faces.

\bibliography{references}
\end{document}